\documentclass[aps,reprint,onecolumn,superscriptaddress,pre]{revtex4-2}
\usepackage{graphicx} % Required for inserting images
\usepackage{amsmath,amssymb}
\usepackage{xcolor}
\usepackage{siunitx}
\newcommand{\E}{\mathbf{E}}
\newcommand{\Ec}{\mathbf{E}^\star}
\newcommand{\Q}{\mathbf{q}}
\newcommand{\R}{\mathbf{r}}
\newcommand{\Eh}{\mathbf{E}_\mathrm{h}}
\newcommand{\Ehc}{\mathbf{E}^\star_\mathrm{h}}
\newcommand{\Ih}{I_\mathrm{h}}
\newcommand{\Es}{\mathbf{E}_\mathrm{s}}
\newcommand{\Esc}{\mathbf{E}^\star_\mathrm{s}}
\newcommand{\Is}{I_\mathrm{s}}
\newcommand{\dt}{\Delta t}

\begin{document}

\title{Ultra-broadband and time-resolved depolarized dynamic light scattering for probing molecular dynamics in supercooled liquids and glasses}

\author{Till Böhmer}
\email{till.boehmer@dlr.de}
\thanks{These authors contributed equally to this work.}
\affiliation{Institute of Frontier Materials on Earth and in Space, German Aerospace Center,
D-51147 Cologne, Germany}
\author{Rolf Zeißler}
\email{rolf.zeissler@pkm.tu-darmstadt.de}
\thanks{These authors contributed equally to this work.}
\affiliation{Institute for Condensed Matter Physics, Technical University of Darmstadt, D-64289 Darmstadt, Germany}
\author{Robin Schwäch}\affiliation{Institute for Condensed Matter Physics, Technical University of Darmstadt, D-64289 Darmstadt, Germany}
\author{Jan P. Gabriel}\affiliation{Institute of Frontier Materials on Earth and in Space, German Aerospace Center, D-51147 Cologne, Germany} 
\author{Florian Pabst}\affiliation{International Centre for Theoretical Physics (ICTP), 34151 Trieste, Italy} 
\author{Thomas Blochowicz}
\email{thomas.blochowicz@physik.tu-darmstadt.de}
\affiliation{Institute for Condensed Matter Physics, Technical University of Darmstadt, D-64289 Darmstadt, Germany}

\date{\today} 
 
\begin{abstract}
\noindent
Dynamic light scattering (DLS) is a versatile technique for probing microscopic dynamics in soft condensed matter. However, applying DLS to supercooled molecular liquids and glasses demands exceptional experimental performance due to weak depolarized scattering, slow relaxation near the glass transition, and the need for quantitative comparison with complementary spectroscopic techniques. In this tutorial we discuss, how a depolarized dynamic light scattering (DDLS) setup can be tailored to meet these challenges. By combining conventional fiber-optical photon correlation spectroscopy, and multispeckle photon correlation imaging with high-frequency DDLS, such a setup allows to capture rotational dynamics across more than 20 orders of magnitude in time. We detail the experimental design required for high signal-to-noise ratios and long-term optical stability, alongside the treatment of coherence and partial heterodyning effects. As we demonstrate, after proper treatment the different detection schemes yield the same electric-field autocorrelation function, enabling the construction of continuous, ultra-broadband DDLS datasets. Furthermore, multispeckle detection enables time-resolved correlation measurements without temporal averaging, extending DDLS to non-equilibrium systems such as aging molecular glasses. This methodology establishes a unified experimental framework for quantitative investigations of equilibrium and non-equilibrium molecular reorientation dynamics over an exceptionally broad time range.
\end{abstract}

\maketitle
%\tableofcontents

\section{Introduction}

In dynamic light scattering (DLS), fluctuations of visible light scattered from a medium are analyzed to extract information about the microscopic motion of scattering centers~\cite{pecora1985dynamic}. DLS is highly versatile because the nature of the scattering centers can vary widely, allowing the study of a broad range of materials, including molecular (supercooled) liquids and glasses~\cite{berne2000dynamic,petzold2010light, schmidtke2012boiling, schmidtke2013reorientational, schmidtke2014relaxation, gabriel2018depolarized, rossler2025relaxation, bohmer2025spectral}, colloidal suspensions or emulsions~\cite{pecora2000dynamic, hassan2015making}, macromolecules such as polymers both in bulk~\cite{patterson1983light, stevens1984dynamic, fytas1993dynamic} and in solution~\cite{chu1985light, fytas1993dynamic}, proteins~\cite{stetefeld2016dynamic}, granular materials~\cite{menon1997diffusing,mayo2025observing,blochowicz2026jamming}, gels~\cite{norisuye2004dynamic, geissler2022dynamic, blanco2000dynamic,pabst2023preserving}, foams~\cite{durian1991multiple, weitz1993diffusing} and many other soft condensed matter systems. While the underlying principle of the technique remains the same across these systems, specific experimental implementations are often tailored to the properties of the particular materials under study. 

This article reports on DLS experiments tailored for the study of rotational dynamics in molecular (supercooled) liquids and glasses covering more than 20 orders of magnitude in time, from femtoseconds to hundred thousands of seconds. This can be achieved by combining data from four different DLS techniques, namely photon-correlation spectroscopy (PCS) employing both fiber-optic and multispeckle detection schemes, as well as tandem Fabry-Perot interferometry and double monochromator-based spectroscopy. The focus of this work lies on the prior two techniques, including the steps that need to be taken to maximize resolution and signal to noise ratio, achieve sufficient long term stability and accurate and stable sample temperature, allowing for the study of rotational dynamics of weakly scattering molecular glass formers at cryogenic temperatures.

We begin by giving a brief introduction to the principle of a typical DLS experiment and an overview of the key research questions that motivated the design and optimization of the experimental setup. Subsequently, the technical foundations of the setup are discussed, followed by an explanation of data acquisition and analysis. Finally, we present and discuss data from two distinct applications, namely time-resolved dynamics during the aging of molecular glasses, and the temperature dependence of molecular dynamics in (supercooled) liquids from the boiling point to the glass transition, as probed by broadband DLS.

\subsection{Brief introduction to DLS}
When a disordered medium is illuminated by a coherent light source, the scattered light forms a characteristic speckle pattern, i.e. a granular pattern of spatial intensity fluctuations perpendicular to the propagation direction of the scattered light. Light and dark speckles arise from constructive or destructive interference of light scattered from different locations within the medium. If the microscopic configuration of the medium evolves in time, the speckle pattern fluctuates, and the temporal variation of the intensity in each speckle reflects the microscopic dynamics. This is the underlying principle of DLS, where the temporal fluctuations of the scattered intensity are probed in order to extract information about the microscopic dynamics of the medium, e.g. the translation of colloidal particles or the rotational dynamics of molecules in (supercooled) liquids and glasses.

In the prototypical PCS experiment, light scattered from a sample is captured by an optical fiber, and the temporal intensity fluctuations $I(t)$ are recorded. The corresponding normalized time-autocorrelation function
\begin{equation}
    \label{equ:g2}
    g_2(\Delta t) = \frac{\bigl\langle I(t)\,I(t+\Delta t) \bigr\rangle_t}{\langle I(t) \rangle^2_t}
\end{equation}
is used to quantify these intensity fluctuations, where $\langle...\rangle_t$ indicates a temporal average, so that $g_2(\Delta t)$ only depends on the lag time $\Delta t$. However, the microscopic dynamics is directly encoded in the electric field time-autocorrelation function 
\begin{equation}
    \label{equ:g1}
    g_1(\Delta t) = \frac{\bigl\langle \Ec(t)\cdot\E(t+\Delta t) \bigr\rangle_t}{\langle |\E(t)|^2 \rangle_t},
\end{equation}
which includes phase information of the scattered light.\\

Assuming the electric field is a complex-valued Gaussian random variable, the Siegert relation~\cite{siegert1943fluctuations} relates $g_1(\Delta t)$ and $g_2(\Delta t)$, which in the simplest case reads as
\begin{equation}
    \label{equ:siegert}
    g_2(\Delta t) = 1+\Lambda g_1(\Delta t)^2.
\end{equation}
Here, $\Lambda$ is the coherence factor, which is unity if the scattered light probed by the detector originates from a single coherence area (single speckle) and satisfies $0<\Lambda<1$ otherwise. More involved versions of Eq. (\ref{equ:siegert}) are discussed below, to account for partial heterodyne contributions, i.e. when the light scattered from the sample is mixed with coherent light from another source, e.g. light reflected from the sample environment.

Depending on the experimental conditions, the decay of $g_1$ reflects different aspects of the dynamics within the sample. A detailed derivation is given, e.g, in the book by Berne and Pecora~\cite{Berne1976}. To summarize briefly, DLS experiments can be performed in different polarization geometries, with the most common ones being vertical-vertical (VV) and vertical-horizontal (VH) geometry. In both geometries the incident light is polarized vertically with respect to the scattering plane, while only a certain polarization component of the scattered light is analyzed, i.e., the vertically polarized one in VV and the horizontally polarized one in VH.

The respective microscopic interpretations of the electric field autocorrelation functions read
\begin{equation}
    g_1^{\mathrm{VV}}(q,\Delta t)=A\,F(q,\Delta t) + B\,F(q,\Delta t)\,C_2(\Delta t)
\end{equation}
and
\begin{equation}
    g_1^{\mathrm{VH}}(q,\Delta t)=D\,F(q,\Delta t)\,C_2(\Delta t).
\end{equation}
Here 
\begin{equation}
    F(q,\Delta t) = \biggl\langle \sum\limits_{j=1}^N \sum\limits_{k=1}^N \exp\left(i \Q\cdot[\R_j(t+\Delta t) - \R_k(t)]\right)\biggr\rangle_t
\end{equation}
is the intermediate scattering function, with the scattering vector $\Q$ and the position of scattering centers $\R_i$.
\begin{equation}
    C_2(\Delta t)=\Bigl\langle P_2\bigl(\cos\bigl[\theta(t,t+\Delta t)\bigr]\bigr) \Bigr\rangle _t
\end{equation}
is the orientational correlation function where $P_2$ is the Legendre polynomial of rank $\ell=2$, $\theta(t,t+\Delta t)$ the change in orientation of the optical polarizability tensor between times $t$ and $t+\Delta t$. $A$,$B$ and $D$ are constants determined by the components of the polarizability tensor of the scattering centers, for detailed expressions see Ref.~\citenum{Berne1976}.

For spherical particles such as colloids, the constants $B$ and $D$ are small compared to $A$ due to their optically isotropic nature. Thus, VV geometry can be used to investigate their translational dynamics. In the case of optically anisotropic molecules, the constants $B$ and $D$ are of similar order as $A$, such that VV and VH geometry both contain contributions from translational and rotational dynamics. However, $F(q,\Delta t)$ is sensitive to translations on the length scale associated with $q$, which is orders of magnitude larger than the size of typical molecules. Therefore, $F(q,\Delta t)$ decays on time scales much longer than those associated with molecular rotations and can be approximated as being unity, which yields
\begin{equation}
    g_1^{\mathrm{VV}}(q,\Delta t)=A + B\,\,C_2(\Delta t),
\end{equation}
\begin{equation}
    g_1^{\mathrm{VH}}(q,\Delta t)=D\,C_2(\Delta t)
\end{equation}
Thus, both geometries give access to the rotational dynamics, while $g_1^{\mathrm{VV}}$ contains a plateau that decays only at very long times. Therefore, to study rotational dynamics in molecular liquids the VH (depolarized) geometry is typically preferred. Such experiments are referred to as \textit{depolarized} dynamic light-scattering (DDLS) and are the focus of the present work.

Besides PCS, other experimental DDLS approaches directly probe the spectral density of the fluctuations of the scattered electric field, i.e., the scattered intensity $I(\nu)$ at different frequency shifts $\nu$ with respect to the frequency of the incident light. Such experiments employ an interferometer or spectrometer to select different spectral components of the scattered light and are usually tunable to scan a specific frequency range. Typical instrumentation involves multipass tandem Fabry-Perot interferometers~\cite{mock1987construction} or different Raman spectrometers~\cite{jones2019raman}. The spectral density of fluctuations of the scattered electric field is related to its autocorrelation function via Fourier-Laplace transform
\begin{equation}
    I(\nu)=\int_{0}^{\infty}g_1(\Delta t)\,e^{-i2\pi\nu \Delta t}d\Delta t\,.
\end{equation}
Therefore, both $g_1(\Delta t)$ and $I(\nu)$ contain equivalent information, however, the corresponding experiments differ in their accessible time- and respective frequency window. While modern PCS experiments can access timescales slower than $\sim 4\,$ns (corresponding to $\nu<40\,$MHz), frequency-domain techniques cover the range from several hundred MHz to the THz regime (corresponding to $1\,\mathrm{ns}>\Delta t>0.01\,$ps). 

By combining these complementary techniques, it is possible to span the entire dynamic range of supercooled liquids, from slow structural relaxation to fast vibrational dynamics. Such a broadband approach to DDLS will be discussed in more detail below. However, we do not provide a detailed description of the high-frequency experimental implementations, as these are well-established techniques. Instead we refer to the works of Cummins et al.~\cite{li1992testing, cummins1996origin}, Rössler et al.\cite{wiedersich2000comprehensive, wiedersich2001light} and Sokolov et al.~\cite{sokolov1995dynamics,surovtsev1998light}. We note that the combination of different DDLS techniques has already been pioneered by Rössler et al.~\cite{petzold2010light, petzold2013evolution, rossler2024glass, rossler2025relaxation}.

\subsection{Motivation and scope of the experimental approach}

The optimization of the DDLS experiments presented in this work was motivated in part by the need to quantitatively compare results for molecular glass formers, for instance monohydroxy or polyhydric alcohols, with other spectroscopic techniques, especially broadband dielectric spectroscopy (BDS). Many of those systems exhibit very weak depolarized scattering due to their low polarizability anisotropy. In addition, a detailed quantitative comparison of spectral features requires transformation of time-domain DDLS data into the frequency domain via Fourier-Laplace transform, which further increases the demands on data quality.

Furthermore, due to the low glass transition temperature of many of these systems (e.g., $T_{\mathrm{g}}\sim100\,$K for 1-propanol) and the strong sensitivity of structural relaxation times to temperature changes, the experiment requires accurate and highly stable temperature control.

Finally, a comprehensive analysis of the temperature-dependence of relaxation dynamics in supercooled liquids requires the possibility of probing slow dynamics up to correlation times of thousands of seconds, i.e., measurement durations of several days. This requires a very high long-time stability of all involved components, such that the intensity autocorrelation functions decay only due to fluctuations in the sample, and not due to changes of the experimental conditions. This is especially difficult considering the requirement to operate at cryogenic temperatures, as most conventional cryostats introduce mechanical vibrations due to vacuum pumps or refrigerators.

The optimized setup described in this work allows for such high-quality measurements over a broad dynamic range and at low temperatures. This enables detailed comparison with BDS and has contributed to disentangling different relaxation processes in dielectric loss spectra, including Debye, structural and secondary relaxations~\cite{gabriel2017debye,gabriel2018depolarized,bohmer2019influence,pabst2019mesoscale,gabriel2020intermolecular,pabst2020dipole,pabst2021generic,bohmer2022glassy,bohmer2023revealing,bohmer2024dipolar,bohmer2025spectral,zeissler2025fresh}.

A second objective has been to enable the study of out-of-equilibrium systems by implementing multi-speckle photon correlation spectroscopy. Analyzing intensity fluctuations within a single speckle, as in conventional DLS experiments, requires extensive time-averaging [compare Eq.~(\ref{equ:g2})]. This requirement imposes significant limitations in order for $g_1(\Delta t)$ to correctly reflect the dynamics of the sample: (i) The dynamics needs to be stationary. Any time-dependence is not considered due to the temporal averaging. (ii) The sample has to be ergodic on the time-scale of the experiment. Ergodicity ensures that temporal averages [Eqs. (\ref{equ:g2}) and  (\ref{equ:g1})] correspond to the ensemble averages. Many systems in soft-matter physics do not fulfill one or both of these requirements. Molecular and colloidal glasses as well as gels are off-equilibrium systems, thus are subject to physical or chemical aging. The dynamics of aging samples is explicitly time-dependent, thus the analysis requires proper two-time correlation functions, i.e., $g_1(t,t+\Delta t)$. In addition, these systems display dynamics on timescales exceeding hours, days or even weeks. Exploring the entire phase space by monitoring a single speckle would require a multiple of this time, implying that, within feasible experimental timescales, glasses and gels usually appear non-ergodic when a single speckle is probed.

Multispeckle DLS solves these issues by monitoring many speckles simultaneously using a camera instead of an optical fiber, thereby replacing the temporal average $\langle...\rangle_t$ by a speckle average $\langle...\rangle_\mathrm{sp}$ and giving access to the two-time intensity autocorrelation function~\cite{viasnoff2002multispeckle,Bissig2003,cipelletti2003time}
\begin{equation}
    \label{equ:g2_multispeckle}
    g_2(t,t+\Delta t) = \frac{\bigl\langle I(t)\,I(t+\Delta t) \bigr\rangle_\mathrm{sp}}{\langle I(t) \rangle_\mathrm{sp}\langle I(t+\Delta t) \rangle_\mathrm{sp}}.
\end{equation}. In previous works, multispeckle DLS has been employed to study the time-evolution of the dynamics, e.g., in aging gels~\cite{Bissig2003,cipelletti2003universal,Ramos2001,Cipelletti2000} and coarsening foams~\cite{Mayer2004,Duri2005,Hoehler2014,Duri2009}.

In this work, we discuss the implementation of a multispeckle approach for probing time-resolved dynamics in molecular glasses. Previously, the setup has been employed to study time-resolved autocorrelation functions of aging molecular and colloidal glasses, providing insight into how aging dynamics is governed by "internal clocks" of the material~\cite{bohmer2024time}.

Multispeckle DDLS can also be applied to stationary systems to probe ultra-slow dynamics. By combining temporal and speckle averaging, a significantly extended time window can be accessed. While fiber-optical DLS provides high temporal resolution at short times, it is limited at long correlation times by the finite correlation window of hardware correlators and the impractically long measurement durations required to access such timescales. In contrast, multispeckle techniques enable reliable measurements at long timescales exceeding hundred thousand seconds~\cite{viasnoff2002multispeckle, cipelletti1999ultralow, cipelletti2003time}, corresponding to very low frequencies (\textmu Hz).

\section{Technical fundamentals of the experimental setup}

This section describes the experimental setup and its individual components, with emphasis on the design principle that enables high-quality DLS measurements, especially at long correlation times. We start by giving an overview of all components, followed by a specific and detailed description of the laser light source, the cryogenic- and sample environment, and the optics employed for multispeckle detection.

As outlined in the previous section, the design of the experiment was governed by three key requirements: (i) High signal-to-noise ratio, enabling measurements on weakly scattering samples and reliable transformation of time domain data into the frequency domain using Fourier-Laplace transformation. (ii) Long-time stability of intensity autocorrelation functions, requiring the suppression of setup-induced scattering-intensity variations due to any mechanical or optical instabilities. (iii) Temperature stability and accuracy to enable the comparison to data of different spectroscopic techniques close to the glass transition temperature, where dynamics is extremely sensitive to small temperature changes.

The basic requirements for conventional fiber-optical DLS setups are well established and described in the literature, e.g. by Pecora~\cite{pecora1985dynamic}. However, their implementation requires particular care for the study of molecular glass formers, where weak scattering signals, long relaxation times, and low temperatures impose constraints. In contrast, the design considerations for multispeckle DLS are less standardized, as the technique is comparatively recent and the applications pursued in this work differ from typical implementations~\cite{cipelletti2003universal, bissig2003intermittent, cipelletti1999ultralow, cipelletti2003time, viasnoff2002multispeckle}. 

\subsection{General description of the experimental setup}

\begin{figure*}[t]
       \centering\includegraphics[width=0.9\textwidth]{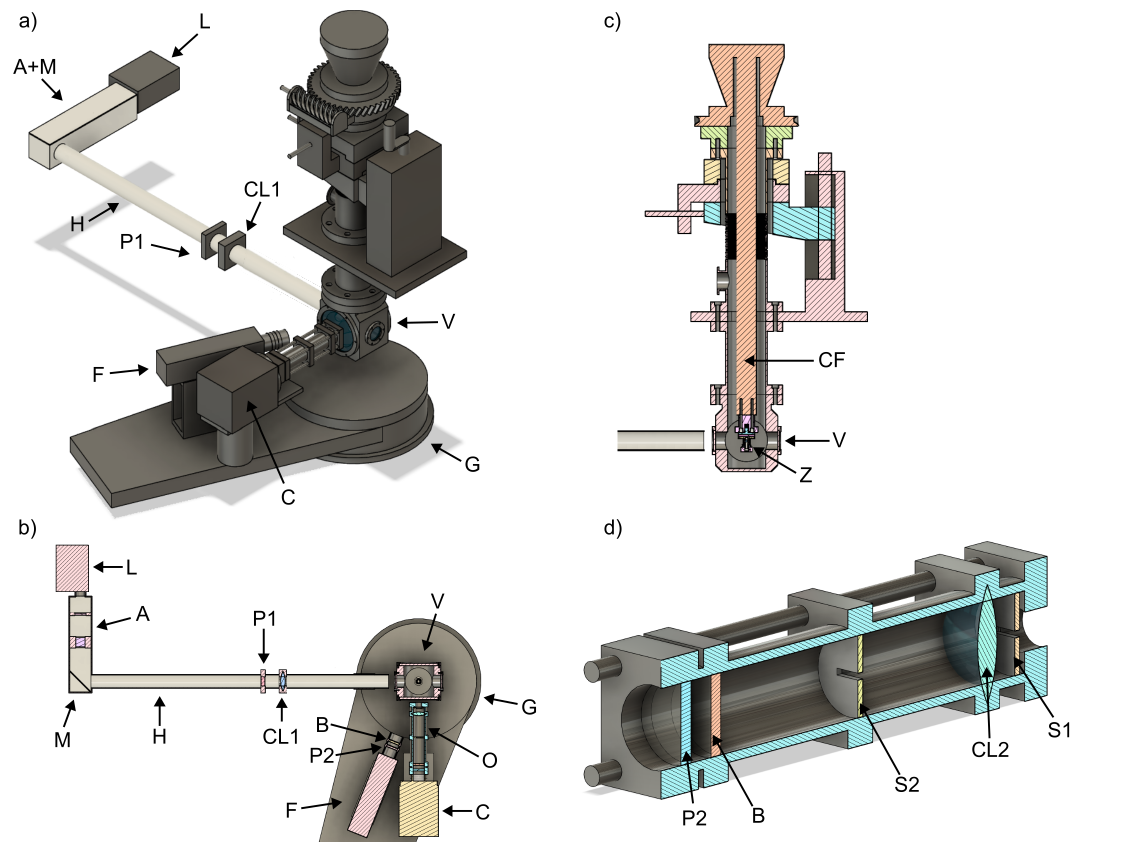}
       \caption{a) Schematic overview of the experimental setup. b) Horizontal cross section of the setup. c) Vertical cross section of the setup. d) Cross section of the multispeckle imaging optics. L: Laser, A: Attenuator, M: Mirror, H: Housing, P1 and P2: Polarizers, CL1 and CL2: Converging lenses, F: Fiber-optics, C: Camera, O: Camera objective, V: Vacuum chamber, G: Goniometer, CF: Cold finger, S1 and S2: Slit apertures, B: Bandpass filter, Z: Sample cell.} 
\label{fig:setup}
\end{figure*}

Fig.\,\ref{fig:setup}a, b and c show a schematic overview of the entire PCS setup. The laser beam (L), after passing through an external variable attenuator (A), is guided towards the sample using an adjustable mirror (M), which ensures that the beam propagates parallel to the scattering plane. The laser beam is contained in a housing (H) for most of its optical path to reduce the amount of airborne dust particles moving through the beam.

A Glan-Laser polarizer (P1) by B.\ Halle with an extinction ratio of $r=10^{-7}$ ensures vertical polarization of the incident beam with respect to the scattering plane. The beam is then focused into the sample using a converging lens (CL1) with a focal length 300\,mm.

The home-built sample cell is mounted on a coldfinger cryostat (CF) inside a vacuum chamber (V). Both, the fiber-optical (F) and the multispeckle detection unit (C) are installed on the goniometer (G), allowing measurements at a scattering angle of 90$^\circ$, respectively. In addition, scattering angles between 20$^\circ-160^\circ$ are accessible, which is advantageous for the investigation of various soft-matter systems. 

In both detection schemes, the scattered light passes a Glan-Thompson polarizer (P2) by B. Halle ($r=10^{-7}$ for the fiber detection, $r=10^{-6}$ with a large aperture of $11.5\times9.8\,\mathrm{mm}^2$ for multispeckle detection) and a 2\,nm bandpass filter (B), before entering the respective detection optics. 

Precise alignment of the beam is achieved using a needle mounted on a high-precision translation stage at a central position on the goniometer arm. When rotating the goniometer from $0^\circ$ to $180^\circ$, the needle translates along the optical axis. For a properly aligned beam, the diffraction pattern formed by the beam irradiating the needle remains symmetric and unchanged between both positions, which can be ensured by adjusting M.
To ensure an optimal alignment of the depolarized polarization geometry, the laser beam is directly coupled into the detector by rotating the goniometer to 180$^\circ$. The beam intensity is strongly attenuated to prevent detector damage. Depolarized geometry is then optimized by minimizing the detected signal through precise rotation of P2 using a high-precision rotation mount.

In case of the fiber detection (F), scattered light is coupled into a single-mode optical fiber with an integrated beam splitter. The transmitted light at both output ports is detected by Count T single-photon counting modules (Laser Components) with low dark-count rate. The use of the beam splitter and two photon counting modules in pseudo-cross detection is standard practice and necessary to reduce afterpulsing, i.e., detection of multiple pulses due to reabsorption of photoelectrons \cite{pecora1985dynamic}. The intensity autocorrelation function is calculated using an ALV-7004/USB-fast hardware correlator, featuring an initial sampling time of 3.125\,ns and a lag time range of 0 to 54\,976\,s. 

To suppress vibrations transmitted through the ground, the entire setup sits on top of a passively damped optical table supported by S2000 laminar flow vibration isolators (Newport).

The following sections discuss in detail several aspects of the setup that require particular attention in order to optimize performance. These include the generation of a highly stable incident beam, the cryogenic sample environment and sample cells, and the multispeckle detection unit.

\subsection{High-stability incident beam}
\label{subsec:laser}

The choice of a suitable light source is crucial for DLS experiments, as it directly determines the signal intensity and any fluctuations of the incident light introduce artifacts in the measured intensity autocorrelation functions, thereby limiting long-time stability. The key requirements are therefore high optical power, low noise, and high mode stability.

The setup employs a Cobolt Samba 05-01 continuous-wave diode pumped frequency doubled Nd:YAG laser (H\"ubner Photonics) operating at a wavelength of 532\,nm with a maximum output power of 500\,mW. 

The beam has a diameter of approximately 700\,\textmu m at the output aperture and a divergence of $<1.2\,$mrad. After focusing with lens CL1, this results in a beam waist of a few hundred microns at the sample position, providing sufficient optical power density for weakly scattering samples such as LiCl-water solutions or alcohols~\cite{zeissler2025fresh,gabriel2017debye,bohmer2022glassy}.

According to the manufacturer, the laser operates in a singe TEM$_{00}$ mode with a linewidth below 1\,MHz, laser-power noise (at 20\,Hz - 20\,MHz) below $0.1$\,\SI{}{\percent}\,rms, and long-term power stability better than 2\,\% over $\pm 2\,$K temperature variation. In practice, optimal stability is achieved by operating the laser at its maximum output power and controlling the incident intensity via the external variable attenuator. 

In addition, the power stability is strongly increased by ensuring a high temperature stability of the laser. This is achieved by using a Cobolt thermoelectric cooling (TEC) plate (H\"ubner Photonics) and by minimizing temperature fluctuations in the laboratory by optimized air-conditioning control. Otherwise, temperature changes in the room can induce laser power fluctuations and respective artifacts in the intensity autocorrelation functions on timescales of minutes up to several hours.

Finally, airborne dust particles interfering with the beam can introduce additional intensity fluctuations. These lead to artifacts in the autocorrelation functions at frequencies around 1-30\,Hz, which become visible after Fourier-Laplace transformation. To suppress these effects, the beam path is enclosed in tubes (H), and the laboratory air is filtered using an air purification system.\cite{gabriel2018depolarisierte}

\subsection{Cryogenic sample environment and sample cells}
\label{subsec:cryo}

\begin{figure}[t]
       \includegraphics[width=0.5\textwidth]{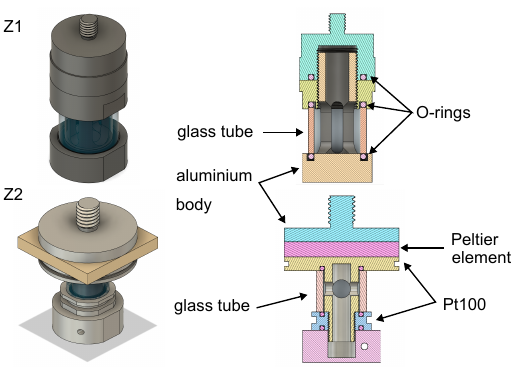}
       \caption{Schematic overview and cross sections of the sample cells. Z1: standard cell for equilibrium measurements at fixed temperatures. Z2: down-scaled version of the standard sample cell equipped with a Peltier element for rapid temperature changes during physical aging experiments.} 
\label{fig:cells}
\end{figure}

Accurate and stable temperature control is essential for studying rotational dynamics in molecular liquids near and below the glass transition, where relaxation times are extremely sensitive to temperature variations. Depending on the fragility of the liquid, a temperature of 1\,K can lead to a change of the structural relaxation time by up to one order of magnitude \cite{cavagna2009supercooled}. Thus, performing measurements in this temperature regime requires special cryogenic environments with high thermal stability.

The sample is contained in a home-built light scattering cell (Z) (see Fig.~\ref{fig:cells}), attached directly to the cold finger of a Konti Spectro A cryostat (Cryovac). The cryostat provides access to temperatures between 4\,K and 450\,K using liquid or gaseous nitrogen or helium. During measurements, gaseous nitrogen is used to maintain the cold finger temperature via a specially designed nitrogen transfer device (Cryovac). Liquid nitrogen is employed only when high cooling rates are required, as its vaporization within the cryostat induces vibrations.

The cold finger is located inside a vacuum chamber (V), which is essential for minimizing thermal gradients and improving temperature accuracy. For optical access, the vacuum chamber is equipped with three anti-reflection coated windows, with a diameter of 3\,cm for beam entry and exit, and of 10\,cm at 90\textdegree to access different scattering angles. To reduce potential contributions of surface reflected light, all aluminum surfaces inside the vacuum chamber are covered by black aluminum foil and the beam dump is located outside of the vacuum chamber.

The cryostat allows for external three-axis translation and one-axis rotation of the cold finger (CF) by a vacuum manipulator, enabling precise control of the position of the scattering volume within the sample. This is convenient for minimizing parasitic heterodyne contributions in the detected light arising from impurities contained in the sample or from reflections of the sample cell.

In contrast to flow cryostats, no exchange gas is used, therefore, insufficient vacuum conditions directly lead to temperature inhomogeneities within the sample cell. To ensure proper vacuum, the vacuum chamber is first evacuated using a turbo-molecular pump. After reaching pressures below $10^{-5}\,$mbar, the turbo-molecular pump (Agilent TwisTorr 74 FS) is isolated from the cryostat to minimize mechanical vibrations. A vibration-free ion-getter pump (Agilent VacIon Plus 75) is used to maintain a vacuum of $10^{-6}\,$mbar and below. Minimizing vibrations is essential for achieving  sufficiently high signal-to-noise ratios in the intensity autocorrelation functions to obtain a satisfactory quality of Fourier-Laplace transformed data.

We use two types of sample cells shown in Fig.\,\ref{fig:cells}, the standard cell for equilibrium measurements at fixed temperatures (Z1) and a down-scaled version equipped with a Peltier element for rapid temperature changes during physical aging experiments (Z2).

The standard cell consists of a quartz-glass tube (outer diameter 20\,mm, wall thickness of 2\,mm) mounted in an aluminum frame. Vacuum sealing is achieved by Viton O-rings pressed on the two polished edges of the glass tube. The middle part of the aluminum body is a tube with four cavities for optical access to ensure the laser to pass trough the sample and the scattered light at 90\textdegree scattering angle to reach the detector. The design ensures that the sample, and especially the scattering volume, is in close contact to the aluminum frame, which improves heat coupling to the cold finger. This reduces temperature gradients and allows for larger cooling rates, which is beneficial for supercooling liquids with a tendency to crystallize. To avoid temperature instabilities due to thermal radiation, the sample cell is enclosed in a gold-plated heat shield, which is directly coupled to the cold finger. To reduce parasitic stray light, all aluminum parts of the sample cell are anodized in matte black. Temperature calibration with respect to absolute temperature units was performed with a Pt-100 temperature sensor placed inside the liquid close to the scattering volume. Temperature stability of the cell is $\pm0.01\,$K, while temperature accuracy with respect to other cryostats in our lab is better than $\pm0.5\,$K

The Peltier-driven cell used for rapid temperature changes was inspired by a design originally created by the Roskilde Glass-and-Time group for use in dielectric experiments \cite{Hecksher2010a}. The concept of the standard cell was scaled down to contain a glass tube with 10\,mm diameter and 1\,mm wall thickness, which is close to the smallest feasible design to not introduce significant additional heterodyne contributions to the detected light. The reduction ensures a lower heat capacity to achieve faster temperature changes. For temperature control, the sample cell introduces a sub-cryostat controlled by a Peltier element (CP402533 by CUI Devices, $25\times25\,\mathrm{mm}^2$). It operates at a fixed cold finger temperature $T_\mathrm{out}$, while the Peltier element controls a temperature offset $\Delta T$ with respect to $T_\mathrm{out}$, leading to a sample temperature $T_\mathrm{in} = T_\mathrm{out}-\Delta T$. $\Delta T$ can be changed much quicker than $T_\mathrm{out}$, therefore the Peltier element allows for more rapid and precise changes in sample temperature.

The Peltier element is glued to the cell using a thin layer of thermally conductive epoxy glue (EP29LPSPAO-1 by Masterbond) and is controlled via the $\pm$10\,V analog output of a Lakeshore 335 temperature controller combined with a Huginn Peltier driver provided by the Roskilde Glass-and-Time group in Denmark~\cite{Jakobsen2019}. Temperature is monitored by a 1.2\,mm$\times$1.6\,mm Pt-100 temperature sensor (IST Innovative Sensor Technology) glued to the cell with thermally conductive epoxy,  while a second identical Pt-100 sits at the bottom of the sample cell to probe temperature gradients within the sample cell.

\subsection{Multi-speckle optics}

Multispeckle light-scattering experiments were introduced in the early nineties~\cite{Wong1993} and have continuously evolved alongside advances in imaging technology. Two principal approaches are commonly employed. In one approach, the camera is positioned in transmission geometry, allowing simultaneous access to a broad range of scattering vectors $\boldsymbol q$.\cite{cipelletti1999ultralow,Cipelletti2000} In the second approach, pursued in this work, a large number of speckles are detected at approximately constant $\boldsymbol q$. Averaging over all detected speckles then allows to obtain time-resolved correlation functions [Eq.~(\ref{equ:g2_multispeckle})]~\cite{viasnoff2002multispeckle,bissig2003intermittent,cipelletti2003time}.

Several imaging geometries can be employed for multispeckle DLS experiments. In the simplest far-field configuration, a single aperture is sufficient to produce a speckle pattern on the camera chip, with light from the entire scattering volume contributing to each detected speckle~\cite{Kirsch1996}. Alternatively, an imaging lens can be used to introduce spatial resolution, such that light detected in a specific region of the speckle pattern originates predominantly from a corresponding subregion of the sample. Such approaches are commonly referred to as \textit{photon correlation imaging} and have been employed, e.g., to detect long-range dynamic heterogeneity in foams and gels~\cite{duri2009resolving}. In recent years, the concept of photon correlation imaging has been adapted and developed for various applications, e.g. to investigate material failure~\cite{Aime2018}, local displacements~\cite{Ju2022} or the yielding transition~\cite{Aime2019,Aime2023} in materials under mechanical strain or shear. In the present work, however, photon-correlation-imaging optics are primarily employed not for spatially resolving heterogeneous dynamics, but to optimize speckle size and detection efficiency. The corresponding design considerations and optical components are discussed in the following.

Considering the goal to probe slow dynamics in deeply supercooled liquids and the aging of molecular glasses, two key requirements can be identified for the multispeckle experiment. First, high temporal resolution is required. In aging molecular glasses, the evolution of the dynamics with waiting time $t$ occurs on timescales comparable to the decay time of the correlation functions with lag time $\Delta t$. Consequently, time-resolved autocorrelation functions must be obtained with minimal temporal averaging, which requires the simultaneous detection of a large number of statistically independent speckles. Second, the setup must allow stable measurements over durations ranging from several hours up to days without producing impractically large datasets. 

These requirements define the central design constraints of the experiment: the speckle size should be sufficiently small to allow many speckles to be recorded within a comparatively small image area, while remaining comparable to or larger than the pixel size of the detector. Otherwise, multiple speckles contribute to a single pixel, reducing the speckle contrast and thus the coherence factor $\Lambda$. For this reason, we employ photon-correlation-imaging optics, which allow controlled adjustment of the speckle size through the positioning of an imaging lens.

A schematic illustration of the optical setup used for the multi-speckle experiment is shown in Fig.~\ref{fig:setup} (d). The scattered light first passes two adjustable slit apertures (S1 and S2), separated by an imaging lens (CL2). The use of slit apertures is motivated by the anisotropic shape of the scattering volume, which corresponds to an elongated cylinder due to the laser beam being observed perpendicular to its propagation direction. The convex lens with focal length $f=$70\,mm is positioned such that the two apertures are located in its focal planes. Adjusting the distance of the lens with respect to the sample and the camera chip allows to control the speckle size at the detector plane.

Subsequently, light passes a horizontally oriented Glan-Thompson polarizer from B. Halle (P2) with extinction ratio $10^{-6}$ and an aperture diameter of 12\,mm, sufficiently large to avoid vignetting. In addition, a 2\,nm bandpass filter (B) is employed to suppress parasitic light from external sources. Finally, the speckle pattern is recorded using a Hamamatsu ORCA-Flash 4.0 V2 SCMOS camera equipped with a 2048\,px x 2048\,px image sensor and individual pixel dimensions of 6.5\,\textmu m $\times$ 6.5\,\textmu m. At full resolution, the camera allows frame rates up to 100\,fps. By restricting to smaller regions of interest, higher frame rates and correspondingly shorter accessible timescales can be achieved. The static and dynamic properties of the recorded speckle patterns are discussed in the following sections.

\section{Multi-speckle correlation analysis and data treatment}

The multispeckle detection scheme introduces additional statistical and optical considerations compared to conventional fiber-optical DLS experiments. In the following, we first characterize the static and dynamic properties of the recorded speckle patterns before discussing the treatment of coherence effects and heterodyne contributions required for reliable extraction of electric-field autocorrelation functions.

\subsection{Static structure of speckle patterns}

\begin{figure*}[t]
       \includegraphics[width=1.0\textwidth]{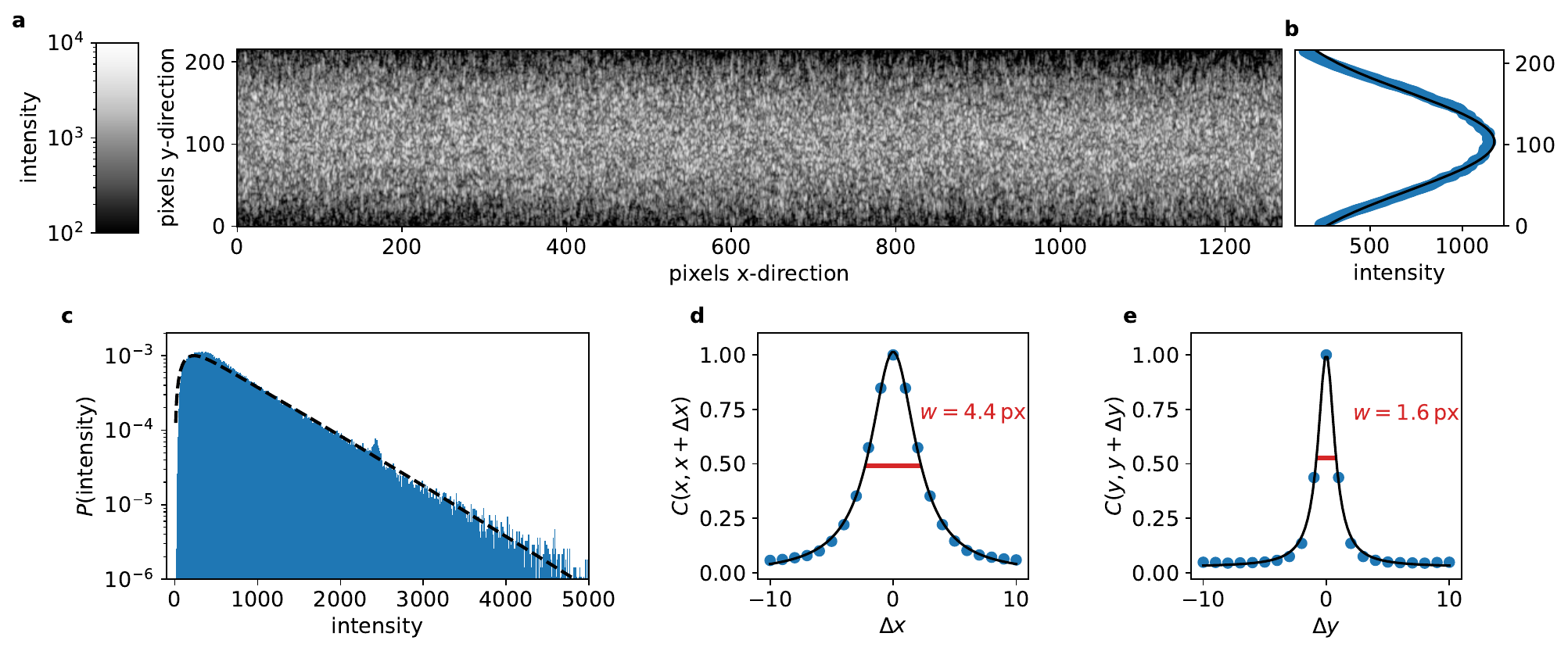}
       \caption{Analysis of the static structure of speckle patterns. a) Example of an intensity speckle pattern of the depolarized scattered light from a supercooled molecular liquid (1-phenyl-1-propanol) slightly above $T_\mathrm{g}$. b) The average intensity profile along the $y$-direction approximately follows a Gaussian shape (black line), reflecting the Gaussian profile of the laser beam, while the average intensity is constant along the $x$-direction. c) Probability distribution of speckle intensities. The experimentally observed distribution corresponds well to Eq.~(\ref{equ:dist}) (dashed line), the theoretical description for a speckle pattern probed using a finite detector resolution. The description via the dashed line includes the effect of the Gaussian intensity profile in the $y$-direction of the image, which yields an additional broadening of the distribution. d) and e) Average spatial intensity autocorrelation function between pixels of distance $\Delta x$ and $\Delta y$. The width of the observed peak is a measure for the speckle size.} 
\label{fig:speckle}
\end{figure*}

The following analysis characterizes the static properties of the recorded speckle patterns and demonstrates that the chosen imaging geometry yields sufficiently small and statistically well-defined speckles for efficient multispeckle detection. 

As an example, a speckle pattern obtained for the horizontally polarized light scattered from the supercooled molecular liquid 1-phenyl-1-propanol slightly above $T_\mathrm{g}$ is shown in Fig.~\ref{fig:speckle}a. Due to the chosen imaging geometry, the camera records a real-space image of the elongated scattering volume produced by the laser beam, resulting in a horizontal stripe with a Gaussian intensity profile along the $y$-direction, as shown in Fig.~\ref{fig:speckle}b. Consequently, only $\sim 6$\% of the full 2048$\times$2048 pixel sensor area is utilized, resulting in comparatively small data volumes even for long-duration measurements.

To analyze the speckle size on the camera chip, the average spatial intensity autocorrelation functions~\cite{viasnoff2002multispeckle} $C(x,x+\Delta x)$ and $C(y,y+\Delta y)$ are evaluated in Fig.~\ref{fig:speckle}d and e. The spatial autocorrelation functions exhibit maxima at $\Delta x=0$ and decay over distances corresponding to the speckle size. The observed correlation profiles can empirically be described by a Lorentzian, from which the full width at half maximum $w$ is determined. Different widths are obtained along the $x$- and $y$-directions, which is a direct consequence of imaging the elongated scattering volume using a spherical lens. Along the $y$-direction, the average speckle size corresponds to $w_y=$1.6\,pixels, i.e. close to the smallest feasible speckle size before substantial reduction of speckle contrast due to pixel averaging occurs. Along the $x$-direction we observe a larger speckle size of $w_x=$4.4\,pixels. Thus, we achieve a near-optimal compromise between maximizing the number of statistically independent speckles and preserving speckle contrast.

The probability distribution of pixel intensities is shown in Fig.~\ref{fig:speckle}c, revealing an asymmetric distribution with a maximum at $I_0>0$. Following the theoretical treatment of speckle statistics by Goodman~\cite{Goodman1975}, the probability density can be described by
\begin{equation}
    \label{equ:dist}
    P(I) = \frac{\left(\frac{m}{I_0}\right)^m I^{m-1}\,\exp\Bigl(-\frac{mI}{I_0}\Bigr)}{\Gamma(m)},
\end{equation}
where the parameter $m$ characterizes the degree of speckle averaging and approaches unity in the limit of fully resolved speckles. The dashed black line in Fig.~\ref{fig:speckle}c represents a fit to the data using $m=2$, additionally considering the Gaussian envelope shown in panel b. The obtained value of $m$ confirms that on average more than one speckle contributes to each pixel, consistent with the speckle size on the order of the pixel size determined from the spatial autocorrelation analysis.
 
\subsection{Speckle dynamics}
The dynamics of the speckle pattern shown in Fig.~\ref{fig:speckle}a is analyzed by calculating the normalized time-resolved intensity time-autocorrelation function $g_2(t,t+\Delta t)$ according to Eq.~\ref{equ:g2_multispeckle}. Due to the symmetry of the recorded speckle pattern, the average intensity is constant along the $x$-direction for fixed $y$. This allows the normalization of the intensity autocorrelations to be performed independently for each row of pixels, yielding
\begin{equation}
\label{trc}
    g_2(t,t+\Delta t) = \frac{n_x}{n_y}\sum\limits_{y=1}^{n_y} \left[  \frac{\sum_{x=1}^{n_x}\,I_{xy}(t)I_{xy}(t+\Delta t)}{\Bigl(\sum_{x=1}^{n_x}\,I_{xy}(t)\Bigr)\Bigl( \sum_{x=1}^{n_x}\,I_{xy}(t+\Delta t)\Bigr)}  \right],
\end{equation}
where $(x,y)$ denote pixel coordinates shown in Fig.~\ref{fig:speckle}a, $I_{xy}(t)$ is the intensity recorded in pixel $(x,y)$ at time $t$, and $n_x$ and $n_y$ denote the number of pixel columns and rows, respectively. The advantage of utilizing the symmetry of the speckle pattern is that no time average is required for obtaining an accurate normalization.

\begin{figure}[t]
       \includegraphics[width=0.49\textwidth]{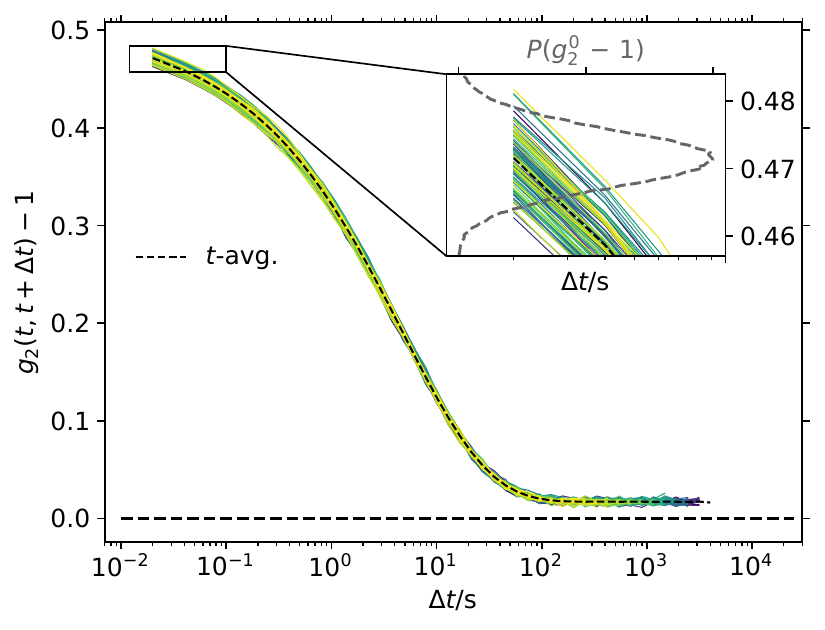}
       \caption{Time-resolved intensity autocorrelation function $g_2(t,t+\Delta t)-1$ obtained for a supercooled liquid sample of 1-phenyl-1-propanol as a function of $\Delta t$. Each color represents a correlation function calculated for a different starting frame at time $t$, while the black dashed line is the average over all starting frames. The inset highlights the slight differences between the different autocorrelation functions at short times and the corresponding distribution $P(g_2^0-1)$ of the short-time autocorrelation values (grey dashed line). The small differences are due to statistical fluctuations associated with the finite number of detected speckles.}
\label{fig:time-resolved-stat}
\end{figure}

Time-resolved intensity autocorrelation functions $g_2(t,t+\Delta t)$ are shown in Fig.~\ref{fig:time-resolved-stat} as a function of lag time $\Delta t$, where different colors correspond to different $t$. Each curve is obtained without any temporal averaging, i.e. each point is obtained simply by correlating two images. All curves collapse within statistical accuracy, reflecting the stationarity of the equilibrium dynamics in the supercooled liquid. Consequently, the average over different $t$, represented by the dashed black line, provides a meaningful description of the equilibrium dynamics.

Small deviations between individual time-resolved autocorrelation arise from statistical fluctuations associated with the finite number of detected speckles. These fluctuations primarily affect the short-time limit
\begin{equation}
    g_2^0(t)=\lim_{\Delta t\to 0}~g_2(t,t+\Delta t),
\end{equation}
which therefore exhibits a finite distribution of values. This is illustrated in the inset of Fig.~\ref{fig:time-resolved-stat}a, showing a magnified view of the short-time regime. The grey line indicates the distribution of $g_2^0$ values, with standard deviation, $\Delta g_2^0=0.004$, corresponding to deviations well below 1\% from the mean value of $g_2^0-1$. 

The small magnitude of these fluctuations demonstrates that the number of statistically independent speckles is sufficiently large to obtain reliable time-resolved autocorrelation functions without the need for additional temporal averaging. Correspondingly, variations of the imaging optics that produce larger average speckle sizes yield larger values of $\Delta g_2^0$, consistent with the reduced number of independent speckles contributing to the ensemble average. 

\subsection{Treatment of coherence and partial heterodyning effects}

\begin{figure*}[t]
       \includegraphics[width=1.0\textwidth]{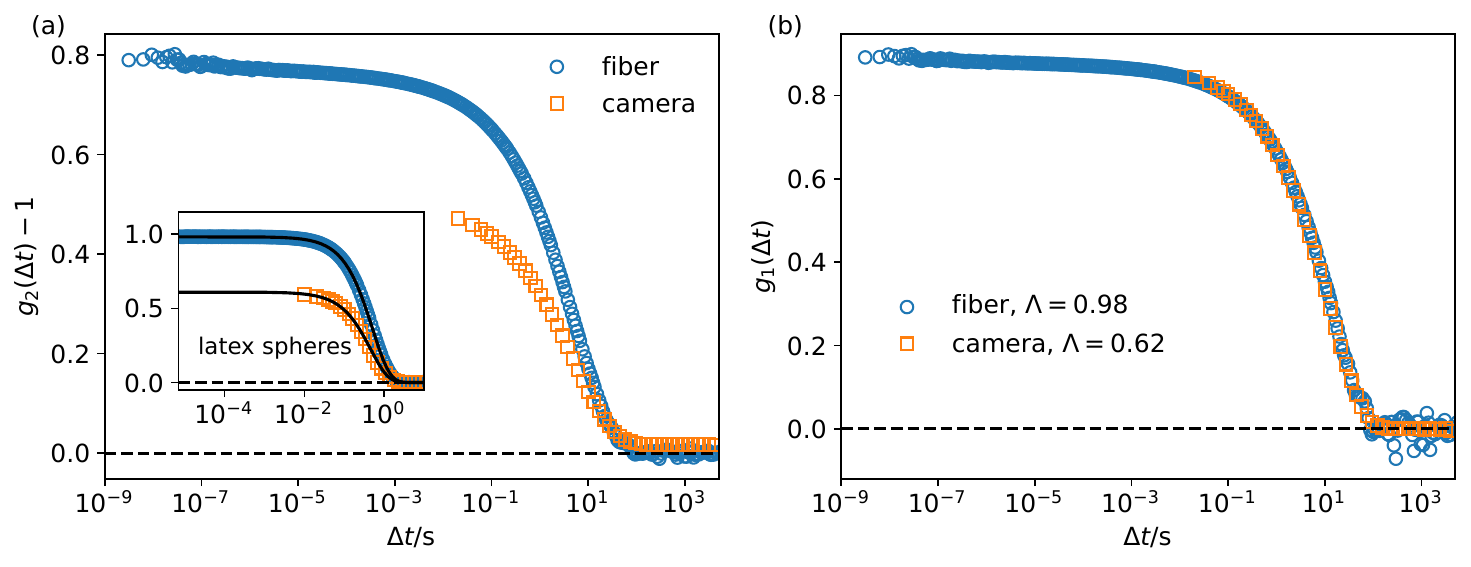}
       \caption{(a) Intensity autocorrelation function as probed by fiber and camera detection for the same sample of supercooled 1-phenyl-1-propanol. Differences occur mainly in the short- and long-time plateau values due to different coherence factors and heterodyning effects. Inset: Corresponding data for latex spheres 5.5,\textmu m polystyrene spheres dispersed in glycerol, the short-time plateau of which gives access to the coherence factors of both techniques, which yields $\Lambda_{\mathrm{fiber}}=0.98$ and $\Lambda_{\mathrm{camera}}=0.62$. (b) Electric field autocorrelation function obtained from the data in (a) by applying the extended Siegert relation [Eq.~(\ref{equ:siegert-ext})]. Besides the different coherence factors, heterodying effects were considered using the procedures discussed below. Camera and fiber data display an almost perfect collapse.}
\label{fig:g2_g1_matching}
\end{figure*}

Both the fiber-optical and the multispeckle DDLS experiments probe the same orientational dynamics and should therefore yield identical electric-field autocorrelation functions. Fig.~\ref{fig:g2_g1_matching}a compares the intensity autocorrelation functions $g_2(\Delta t)$ obtained with the two detection schemes for the same sample of supercooled 1-phenyl-1-propanol. Here and throughout the following discussion, quantities expressed as functions of the lag time $\Delta t$ alone denote $t$-averaged quantities. The comparison shows that the multispeckle data exhibit both a reduced short-time amplitude and a long-time plateau above unity relative to the fiber-optical data. These differences arise from the different degrees of speckle averaging and from the distinct ways in which heterodyne light contributes to the measured intensity autocorrelation functions in the two detection schemes.

Quantitatively, both effects are described by the generalized Siegert relation for partially heterodyne detection~\cite{pabst2017molecular,Bremer1993},
\begin{equation}
    \label{equ:siegert-ext}
    g_2(\Delta t) = 1+\Lambda C^2\, g_1(\Delta t)^2 + 2\Lambda C(C-1)\,g_1(\Delta t),
\end{equation}
where the coherence factor $\Lambda$ quantifies the reduction of speckle contrast due to the resolution of the detector and
\begin{equation}
\label{equ:def_C}
C=\frac{\langle I_s\rangle}{\langle I\rangle}
\end{equation}
quantifies the degree of heterodyning. Here, $\langle I_s\rangle$ denotes the average intensity scattered from the sample and $\langle I\rangle$ the total detected intensity. The homodyne limit corresponds to $C=1$, whereas $0<C<1$ indicates partial heterodyning due to static light contributions that do not reflect the sample dynamics.

Inspection of Eq.~(\ref{equ:siegert-ext}) reveals that both $\Lambda$ and $C$ influence the measured intensity autocorrelation function. The reduced short-time amplitude of the multispeckle data is primarily a consequence of its smaller coherence factor $\Lambda$ compared to the fiber-optical experiment. Heterodyne contributions, described by the parameter $C$, affect the two detection schemes differently. In the multispeckle experiment they manifest through an increased long-time plateau, whereas in the fiber-optical experiment they primarily reduce the short-time amplitude while leaving the long-time limit unchanged at unity~\cite{Xue1992,Joosten1991}. These differences are discussed in detail below. Both $\Lambda$ and $C$ must therefore be determined before the electric-field autocorrelation functions obtained from the two detection schemes can be compared or merged to produce a broadband data set.

To determine the coherence factors of the two experiments, measurements were performed on a dilute suspension of 5.5\,\textmu m polystyrene spheres dispersed in glycerol in polarized geometry. For such a system, heterodyne contributions can be neglected due to the large scattering intensity from the sample and the generalized Siegert relation reduces to the conventional homodyne form, Eq.~(\ref{equ:siegert}). Thus, $\Lambda$ can be determined by extracting the short-time plateau from these data. A corresponding fit yields $\Lambda_{\mathrm{fiber}}=0.98$ and $\Lambda_{\mathrm{camera}}=0.62$ for the fiber-optical and the multispeckle experiment, respectively.

The reduced coherence factor of the multispeckle setup is consistent with the static characterization of the speckle pattern discussed above. In particular, the intensity distribution shown in Fig.~\ref{fig:speckle}c is described by $m\approx2$, indicating partial spatial averaging. For the Gamma-distributed intensity statistics of Eq.~(\ref{equ:dist}), the normalized intensity variance is given by $1/m$, yielding $1/m\approx0.5$. Since the coherence factor $\Lambda$ is directly related to the normalized variance of speckle intensities, a value of $\Lambda$ of similar magnitude is expected. The reasonable agreement with the independently determined value $\Lambda_{\mathrm{camera}}=0.62$ therefore provides an internal consistency check.

Applying Eq.~(\ref{equ:siegert-ext}) using the experimentally determined coherence factors $\Lambda_{\mathrm{fiber}}=0.98$ and $\Lambda_{\mathrm{camera}}=0.62$ yields the electric-field autocorrelation functions shown in Fig.~\ref{fig:g2_g1_matching}b. For the present sample, heterodyne contributions are weak and only require a minor correction. Nevertheless, they were accounted for using the procedure described in detail below. After correction, the fiber-optical and multispeckle data collapse onto a single curve, demonstrating that both experiments probe the same underlying molecular dynamics.

While heterodyne contributions are comparatively small in the example shown here, they become important for even weaker scattering samples. Reliable determination of the heterodyne parameter $C$ is therefore essential. In the following, corresponding approaches for multispeckle and fiber-optical data are discussed separately. The need for separate approaches originates from a fundamental difference between temporal averaging in a single speckle and spatial averaging over many speckles. In conventional fiber-optical DLS, the intensity autocorrelation function is obtained by temporal averaging within a single speckle. Consequently, the intensity of the static heterodyne contributions remains constant throughout the experiment and only affects the short-time amplitude of the correlation function. A procedure for extracting the degree of heterodyning from the short-time amplitude of fiber data is presented below.
By contrast, multispeckle DLS relies on averaging over a large number of speckles. Since also the static light contribution forms a spatially varying speckle pattern on the detector, speckle averaging yields a corresponding long-time plateau above unity in the intensity autocorrelation function, from which the degree of heterodyning can be determined directly.

\subsubsection{Approach for multi-speckle data}
To quantitatively relate the long-time plateau of the multispeckle intensity autocorrelation function to the heterodyne parameter $C$, we consider the influence of static heterodyne light on the speckle-averaged intensity correlations. For this purpose, the electric field detected on the camera sensor is decomposed into a fluctuating contribution originating from the sample dynamics $I_\mathrm{s}$ and a static contribution $I_\mathrm{h}$ representing the heterodyne light.

We assume that the electric field $\E(\R,t)$ detected on the camera chip surface $\Omega$ consists of a dynamically fluctuating contribution $\Es(\R,t)$, originating from light scattered by the sample, and a static contribution $\Eh(\R,t)$, representing the heterodyne light, such that
\begin{equation}
    \E(\R,t) = \Es(\R,t) + \Eh(\R,t).
\end{equation}
At a fixed point $\R\in\Omega$ on the camera chip, the intensity $\Is = |\Es\cdot\Esc|$ fluctuates in time, reflecting the sample dynamics. By contrast, $\Ih = |\Eh\cdot\Ehc|$ is time-independent at fixed $\R$, but varies across the detector surface, i.e., as a function of $\R$, therefore forming a speckle pattern. As a consequence, the normalized intensity autocorrelation function obtained by averaging over time at fixed $\R$ (moving time average), is not identical to the one obtained by averaging over $\R\in\Omega$ (speckle average).

The intensity detected at the camera is given by
\begin{equation}
    I(\R,t) = |\E(\R,t)\cdot\Ec(\R,t)| = \bigl| \Es(\R,t)+\Eh(\R,t)\bigr|\cdot\bigl| \Esc(\R,t)+\Ehc(\R,)\bigr|.
\end{equation}
The corresponding non-normalized intensity autocorrelation function reads
\begin{align}
    \label{equ:G2_hetero}
    G_2(t,t+\Delta t) =& \Bigl\langle \bigl[\bigl( \Es(t)+\Eh(t)\bigr)\bigl( \Esc(t)+\Ehc(t)\bigr)\bigr]\cdot \bigl[\bigl( \Es(t+\dt)+\Eh(t+\dt)\bigr)\bigl( \Esc(t+\dt)+\Ehc(t+\dt)\bigr)\bigr] \Bigr\rangle,
\end{align}
where the spatial coordinate $\R$ has been omitted in the last line for readability.

Expanding Eq.~(\ref{equ:G2_hetero}) yields sixteen terms, several of which vanish or can be simplified using the following properties: (i) $\Es$ and $\Eh$ are statistically independent, such that their products average to zero. $\Is$ and $\Ih$ are statistically independent and their averages can be factorized. (ii) Only products involving complex-conjugate field pairs yield non-zero ensemble averages. (iii) Ensemble-averaged intensities are stationary. (iv) The static contribution $\Es$ is time-independent. Using these relations, the non-normalized autocorrelation function reduces to
\begin{align}
G_2(t,\dt) &= \bigl\langle \Is(t)\Is(t+\dt) \bigr\rangle + 2\bigl\langle\Is \bigr\rangle \bigl\langle\Ih \bigr\rangle + 2 \langle \Ih\rangle \bigl\langle\Es(t)\cdot\Esc(t+\dt) \bigr\rangle +  \bigl\langle \Ih^2 \bigr\rangle.
\end{align}
In the long-time limit, $\dt$ is much larger than the relaxation time of the sample, such that the scattered fields and intensities at times $t$ and $t+\dt$ become uncorrelated. In this limit, the normalized intensity autocorrelation function, therefore, simplifies to
\begin{align}
\label{equ:limit_g2}
    \lim\limits_{\dt\to\infty}g_2(t,t+\dt)\equiv g_2^\infty &= \frac{\bigl\langle \Is \bigr\rangle^2 + 2\bigl\langle\Is \bigr\rangle \bigl\langle\Ih \bigr\rangle +  \bigl\langle \Ih^2 \bigr\rangle}{\langle \Is + \Ih \rangle^2} = 1 + \frac{\bigl\langle \Ih^2 \bigr\rangle-\bigl\langle \Ih \bigr\rangle^2}{\langle I \rangle^2},
\end{align}

Eq.~(\ref{equ:limit_g2}) highlights the fundamental difference between temporal averaging within a single speckle and spatial averaging over many speckles. For a single speckle, $\Ih$ is constant and therefore $\langle\Ih^2\rangle=\langle\Ih\rangle^2$, yielding $g_2^\infty=1$. In contrast, multispeckle experiments average over speckles with different values of $\Ih$. Consequently, the variance of the scattered intensity contributes to the long-time limit and produces a plateau above unity.

Assuming, in the spirit of the Siegert relation, that $\Eh(\R,t)$ is a complex-valued Gaussian random variable, its fourth moment can be expressed in terms of products of second moments~\cite{Isserlis1918}. In addition, finite detector resolution must be considered, since individual pixels probe more than a single speckle, which yields~\cite{berne2000dynamic}
\begin{equation}
    \bigl\langle \Ih^2 \bigr\rangle = (1+\Lambda)\langle \Ih\rangle^2,
\end{equation}
where it was assumed that the coherence factor $\Lambda$ is identical for the scattered and the static electric-field contributions, due to being probed using the same optics.

Substituting into Eq.~(\ref{equ:limit_g2}) gives
\begin{align}
     g_2^\infty = 1 + \Lambda\frac{\langle \Ih\rangle^2}{\langle I\rangle^2} = 1 + \Lambda\left(1-\frac{\langle \Is\rangle}{\langle I \rangle}\right)^2.
\end{align}
This implies that the degree of heterodyning $C$ can be extracted directly from the long-time plateau of speckle-averaged intensity time-autocorrelation functions as
\begin{equation}
    C = \frac{\langle\Is\rangle}{\langle I\rangle} = 1-\sqrt{\frac{g_2^\infty-1}{\Lambda}}.
    \label{equ:C_multi}
\end{equation}

\subsubsection{Approach for fiber-optical data}
For data obtained by monitoring the scattered intensity using an optical fiber, the degree of heterodyning can be determined from the value of the short-time limit $g_2^0$ of the intensity autocorrelation function, which is reduced compared to the homodyne scenario. However, this procedure requires knowledge of the amplitude $A_\mathrm{fast}$ of fast relaxation contributions that remain unresolved by the hardware correlator, because they occur on timescales shorter than the smallest accessible correlation time $\Delta t_0=3.125\,$ns. In molecular liquids, these unresolved contributions mainly arise from intra- and intermolecular vibrations.

Since these contributions are not resolved in the fiber-optical experiment, the maximum short-time value of $g_1(\Delta t)$ is reduced from unity to $1-A_\mathrm{fast}\equiv\lambda$, even in a scenario where $C=1$ and $\Lambda=1$. Determining $A_\mathrm{fast}$ is possible by means of high-frequency DDLS experiments though, such that $C$ for the fiber-optical experiment can be calculated by solving the extended Siegert relation Eq.~(\ref{equ:siegert-ext}) under the condition $g_1(t,t+\dt_0)=\lambda$~\cite{pabst2017molecular}. This yields a quadratic equation with solutions~\footnote{We note that in a prior publication~\cite{pabst2017molecular}, we only discussed $C_-$ as a possible solution.}
\begin{equation}
    C_\pm = \frac{1\pm \sqrt{1+g_2^0\bigr(\Lambda^{-1}-2(\lambda\Lambda)^{-1}\bigr)}}{2-\lambda}.
    \label{equ:hetero_siegert}
\end{equation}
There is no direct way to determine which of the two solutions is the right one. A lower constraint can, however, be estimated via the assumption that the multispeckle experiment will always detect a larger heterodyne contribution than the fiber-optical one due to its  larger acceptance angle and, therefore, $C_{\mathrm{fiber}}\geq C_\mathrm{camera}$. Hence, if $C_-<C_\mathrm{camera}$, by default we use $C_+$. 

Thus, by combining high-frequency DDLS to determine $\lambda$ and multispeckle DDLS for choosing the correct solution of Eq.~(\ref{equ:hetero_siegert}), Eq.~(\ref{equ:siegert-ext}) can be used to obtain the electric field autocorrelation function $g_1(\Delta t)$ from $g_2(\Delta t)$ for fiber-optical data. The resulting transformation can be cross-validated by verifying that the electric-field autocorrelation functions obtained from the three DDLS experiments are mutually consistent. This procedure is demonstrated below for the molecular glass former diethyl phthalate.

\subsection{High frequency DDLS or spectral density measurements}
\begin{figure*}[tbh]
       \includegraphics[width=0.9\textwidth]{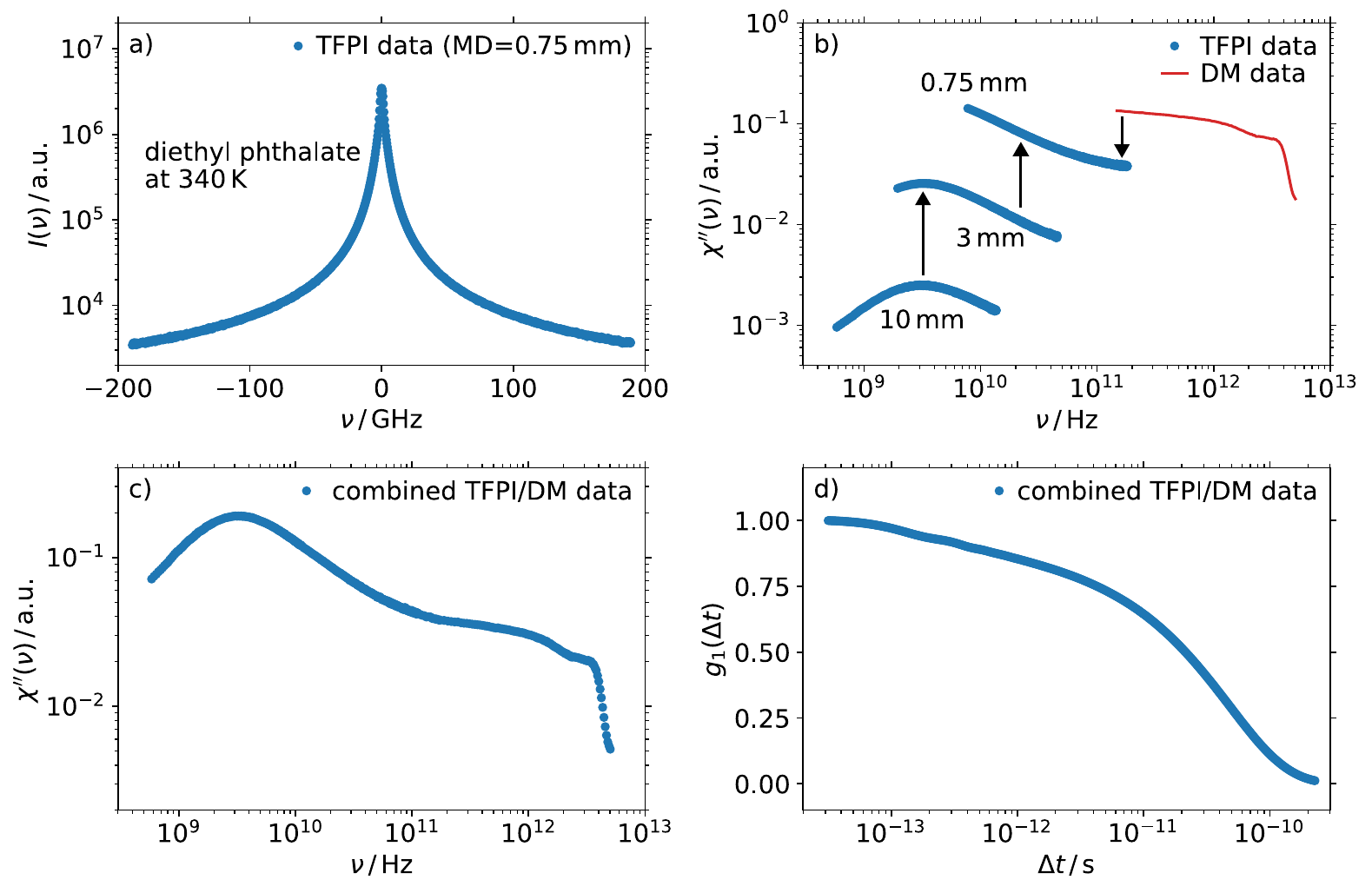}
       \caption{Data transformation procedure for spectral density measurements. a) Raw spectral density data for diethyl phthalate at 340\,K obtained by tandem Fabry-Perot interferometry (TFPI) at a mirror spacing of 0.75\,mm. b) Susceptibilities obtained by transformation of spectral density data via Eq.~\ref{eq:fdt} at different mirror spacings of the TFPI and from double monochromator (DM) measurements. c) Final susceptibility spectrum from shifting the datasets presented in b) to achieve overlap. d) Field autocorrelation function $g_1(\Delta t)$ obtained by inverse Fourier-Laplace transformation of the susceptibility data.} 
\label{fig:spectral_density}
\end{figure*}

Fig.\,\ref{fig:spectral_density} shows the data transformation procedure for high frequency DDLS on the example of diethyl phthalate at 340\,K, i.e., spectral density measurements using a multipass tandem Fabry-Perot interferometer (TFPI) by JRS scientific instruments (model TFP-1) and a Jobin Yvon U1000 double monochromator (DM). In both experiments the spectral density, i.e., the scattered intensity at different frequency shifts relative to the laser frequency, is accessed. The depolarized component of the scattered light is again selected by using two polarizers. The imaginary part of the DDLS susceptibility $\chi^{\prime\prime}(\nu)$ is then calculated via the fluctuation dissipation theorem:
\begin{equation}
\chi^{\prime\prime}(\nu)=\frac{I(\nu)}{n(\nu,T)+1}
\label{eq:fdt}
\end{equation}
where $n(\nu,T)=(\mathrm{exp}(h\nu/k_{\mathrm{B}}T)-1)^{-1}$ is the Bose temperature factor. Fig.\,\ref{fig:spectral_density} a) shows the spectral density measured by the TFPI at a mirror spacing of 0.75\,mm. The mirror distance determines the resolution of the interferometer and the free spectral range, i.e., the frequency range which can be scanned without contributions from neighbouring transmission maxima of the TFPI. To access a broad frequency range, measurements at several mirror spacings have to be performed, as demonstrated in part b) of the figure. To extend the data to frequencies of THz frequencies the TFPI data is combined with DM data. All datasets are combined by shifting them with constant factors to achieve overlap and then cutting the datasets in favor of the smaller mirror spacings, which usually have a larger signal-to-noise ratio due to loss of intensity in the cavities scaling with mirror spacing. To obtain correct relative values of the susceptibility between measurements of a sample at different temperatures or measurements of different samples, the same mirror spacing (here 0.75\,mm) needs to remain unshifted and serves as a reference. Fig.\,\ref{fig:spectral_density} c) shows the combined TFPI/DM dataset. The susceptibility can then be transformed into a field autocorrelation function $g_1(\Delta t)$ by inverse Fourier-Laplace transformation.
As explained above $g_1(\Delta t)$ obtained by the spectral density measurements can then be used for correct transformation of the fiber-optic and multispeckle photon correlation data. This will be demonstrated below.

\subsection{Combining data sets from different DDLS experiments}
 \begin{figure*}[tbh]
       \includegraphics[width=0.5\textwidth]{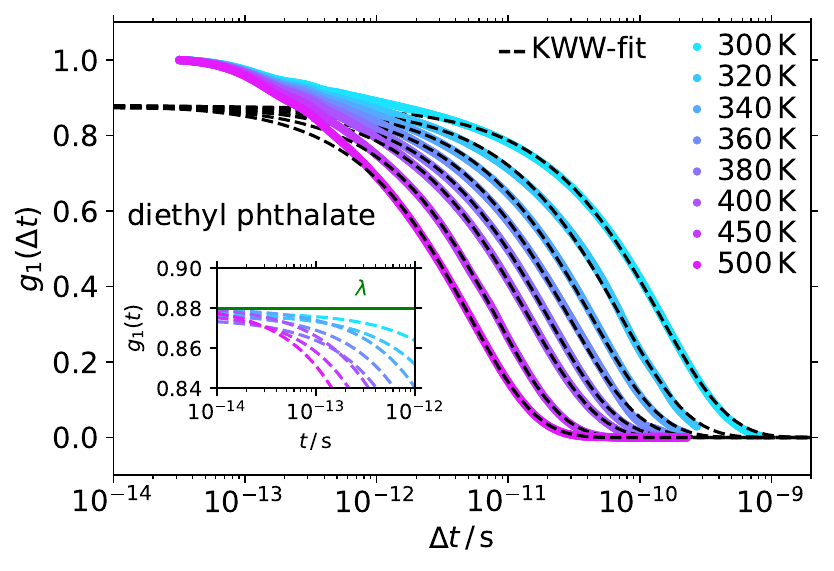}
       \caption{a) Field autocorrelation functions $g_1(\Delta t)$ from combined tandem Fabry-Perot (TFPI) and double monochromator (DM) measurements of diethyl phthalate covering a large temperature range from slightly above the melting point to close to the boiling point. Dashed curves are fits by the KWW function to the second step in the correlations function, i.e., fit range $\Delta t>10^{-12}\,$s. The inset shows the short time limit of the KWW functions employing the same color code as the data presented in the main figure. The solid green line displays the average short time plateau value $\lambda$ of the KWW functions.} 
\label{fig:TFPI_timedomain}
\end{figure*}
\begin{figure*}[tbh]
       \includegraphics[width=\textwidth]{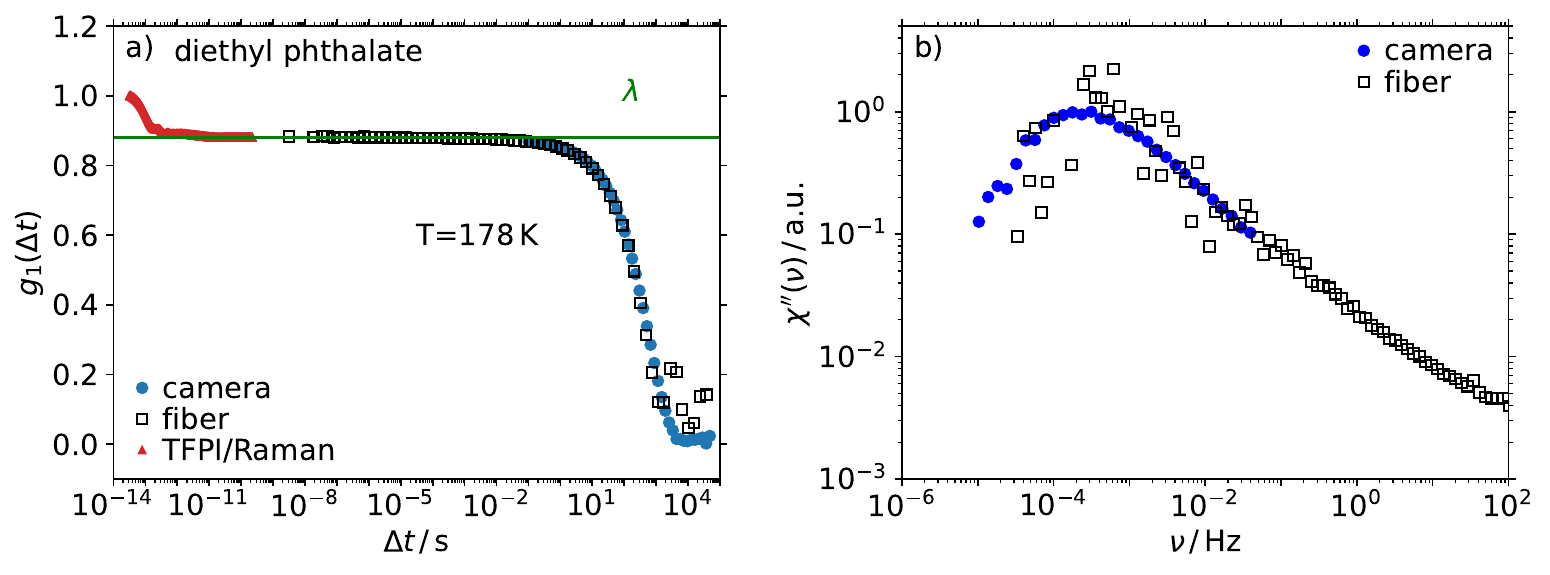}
       \caption{Demonstration of the procedure combining the data obtained by the different techniques in a) the time domain and
b) the frequency domain. The overlap between camera and fiber data is used to confirm the matching of the two techniques and to check that the raw data were treated correctly. To produce the final combined datasets the fiber data overlapping the camera data are removed.} 
\label{fig:camera_fiber_combination}
\end{figure*}

Fig.\,\ref{fig:TFPI_timedomain} shows electric-field autocorrelation functions $g_1(\Delta t)$ obtained by combination of TFPI and DM measurements as described in the previous section. The presented data cover a temperature range of 200\,K, spanning a large part of the temperature range from the melting point $T_{\mathrm{m}}=270\,$K to the boiling point $T_{\mathrm{b}}=575\,$K. The autocorrelation functions show a clearly resolved two step decay, containing both the fast microscopic as well as the slower relaxational decay. Due to its weak temperature dependence, the microscopic correlation decay remains at timescales around ps even when the liquid is strongly supercooled, thus it is unresolved by the PCS techniques discussed above. 

The high-frequency data shown in Fig.\,\ref{fig:TFPI_timedomain} allow to determine the amplitude of the short-time dynamics $A_\mathrm{fast}=1-\lambda$. One way to extract this information is to fit $g_1(\Delta t)$ by a KWW function, while limiting the fit range to the slow decay of the correlation function, as demonstrated in Fig.\,\ref{fig:TFPI_timedomain} by the dashed black lines. Here, the KWW function is merely used to extrapolate the intermediate plateau and is not intended as a physical description of the relaxation process. Another possible approach is to fit the full correlation decay by some more elaborate fit model, which, however, relies on assumptions regarding the nature of the microscopic correlation decay. The inset of Fig.\,\ref{fig:TFPI_timedomain} shows a zoom to the plateau of the KWW fit functions, confirming that the intermediate plateau of $\lambda=0.88(1)$ is virtually temperature independent over a temperature range of 200\,K. It is therefore reasonable to extrapolate this value to lower temperatures and to use it for the transformation and normalization of $g_1(\Delta t)$ obtained from the fiber-optical experiment.

Fig.\,\ref{fig:camera_fiber_combination} shows how data from the different DDLS techniques can be combined to form broadband datasets in the a) time domain and b) frequency domain for the example of diethyl phthalate at 178\,K. For that purpose, all datasets are first transformed independently into electric-field autocorrelation functions $g_1(\Delta t)$. 

For the multispeckle data, evaluation of the long-time plateau according to Eq.~(\ref{equ:C_multi}) yields a degree of heterodyning of $C_\mathrm{camera}=0.97$. Together with the independently determined value $\lambda=0.88$, only the solution $C_+>C_\mathrm{camera}$, thus ruling out the solution with negative sign $C_-$ as a solution of Eq.~(\ref{equ:hetero_siegert}) and resulting in $C_\mathrm{fiber}\approx1$. With all parameters of the generalized Siegert relation now determined independently, the fiber-optical and multispeckle intensity autocorrelation functions can be transformed into the corresponding electric-field autocorrelation functions.

Looking at the combined data set Fig.\,\ref{fig:camera_fiber_combination}, we observe a sufficient correspondence between the three data sets, which represents a cross-validation of the discussed procedures of obtaining $g_1(\Delta t)$. At long lag times, the advantage of the multispeckle setup becomes evident through its substantially improved signal-to-noise ratio compared to the fiber-optical data. Both measurements were taken over approximately one day, which for the fiber-optical DDLS setup is insufficient to produce a smooth correlation function at correlation times $\Delta t\gtrsim 10^3\,$s. By contrast, the multispeckle experiment only requires measurement times on the order of the desired correlation time to produce reliable values of the correlation function. The combination of the different techniques therefore enables the construction of continuous ultra-broadband DDLS datasets.

\section{Discussion and Conclusions}
Having established the methodology and validated the consistency of the different DDLS techniques, we conclude by highlighting two applications that particularly benefit from the improved capabilities of the setup. 

\begin{figure*}[t]
       \includegraphics[width=\textwidth]{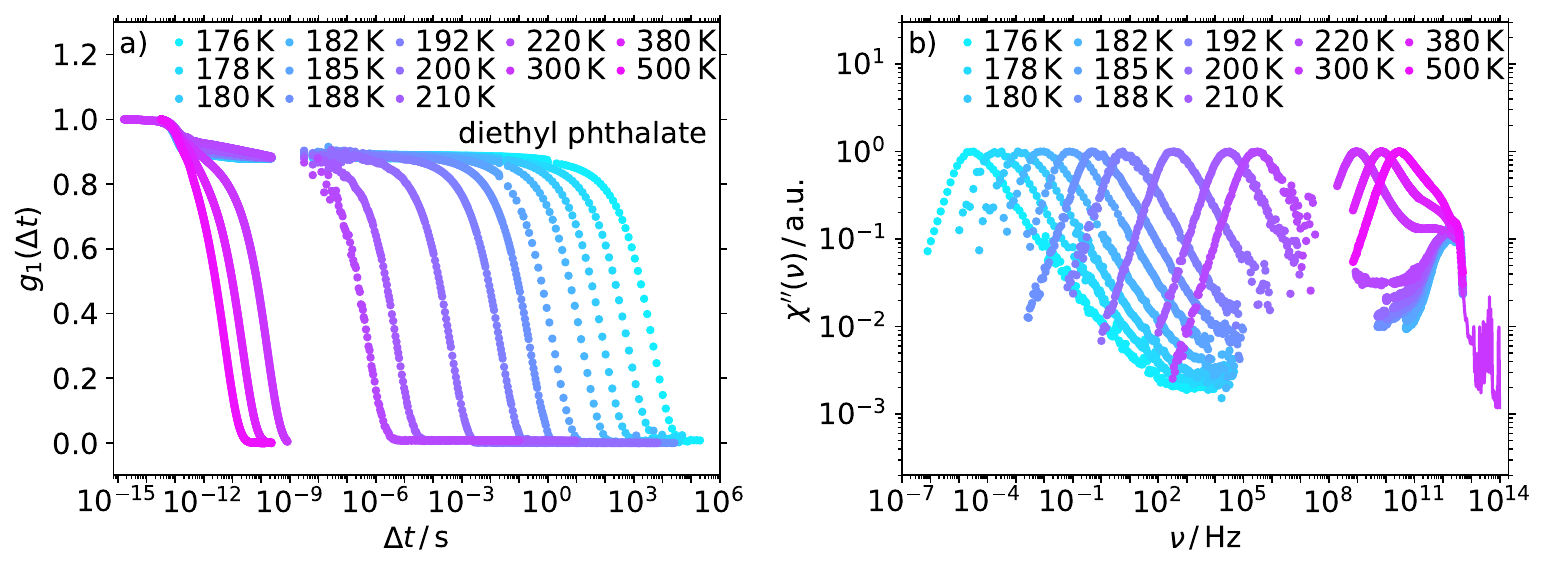}
       \caption{Ultra-broadband DDLS datasets for diethyl phthalate in the a) time domain and b) frequency domain. The dataset covers temperatures
ranging from below the glass transition temperature of Tg$\sim$180K to far above the melting point Tm = 270 K.} 
\label{fig:broadband_DDLS}
\end{figure*}

\textbf{Ultra-broadband DDLS.}
The combination of the different DDLS techniques presented above enables the construction of ultra-broadband datasets covering an exceptionally large dynamic range in both the time and frequency domains. Fig.\,\ref{fig:broadband_DDLS} shows broadband DDLS datasets for the molecular glass former diethyl phthalate in the (a) time domain and (b) frequency domain. The datasets span temperatures from well above the melting point to below the conventional definition of the glass transition temperature ($\tau\approx 100\,\mathrm{s}$). At the lowest temperature the sample was equilibrated for several days to ensure thermal equilibrium. The presented dynamic range covers timescales shorter than ps up to $2\cdot10^{5}\,\mathrm{s}$, i.e., frequencies of sub-\textmu Hz to THz. In fact, the time and frequency range can be further extended at the high frequency side up to $10^{14}\,$Hz into the Raman bands, as is demonstrated for the dataset at 300\,K, and is limited at the low frequency side only by equilibration time and measurement duration, as shown in Fig.\,\ref{fig:broadband_DDLS}. Thus, apart from the small gap around $10 - 100\,$MHz, it is possible to cover the full range of relaxation dynamics in supercooled liquids, comparable to the one reported for broadband dielectric spectroscopy~\cite{lunkenheimer2000glassy, kremer2002broadband, lunkenheimer2018glassy}.

\begin{figure}[t]
       \includegraphics[width=0.49\textwidth]{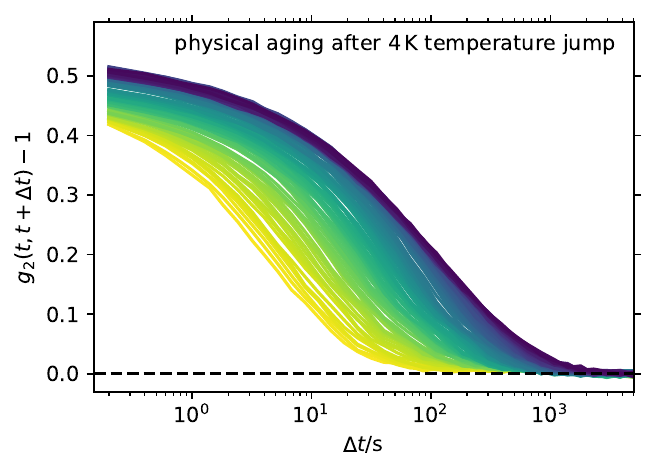}
       \caption{Time-resolved intensity autocorrelation functions of a non-stationary physically aging sample of 1-phenyl-1-propanol, which was exposed to a 4\,K temperature down-jump at $t=0$. Different aging times $t$ are indicated by the colors changing from light to dark with increasing $t$. Each curve is obtained without temporal averaging.}
\label{fig:aging}
\end{figure}

\textbf{Physical aging.}
The multispeckle experiment provides direct access to time-resolved intensity autocorrelation functions without the need for temporal averaging, thereby enabling the investigation of non-stationary processes that are inaccessible to conventional single-speckle PCS. Fig.~\ref{fig:aging} shows data for the physical aging of a glass made from 1-phenyl-1-propanol by exposing the material to a 4\,K temperature down-jump. After the down-jump, the time-resolved intensity autocorrelations reveal a continuous slow-down of the dynamics until eventually reaching a new stationary equilibrium state. 
Beyond providing a direct visualization of the slowing dynamics during aging, these measurements enable a quantitative analysis of non-equilibrium relaxation. In particular, it was recently shown that the time-resolved autocorrelation functions obtained with the present setup allow direct determination of the material time of physical aging~\cite{Narayanaswamy1971,Dyre2015}, thereby establishing a direct connection between equilibrium and non-equilibrium dynamics in molecular glasses~\cite{bohmer2024time}.

In this tutorial, we have used our experimental setup as a practical example to discuss how depolarized dynamic light-scattering experiments can be optimized for the investigation of molecular supercooled liquids and glasses. The setup combines conventional fiber-optical photon-correlation spectroscopy and multispeckle photon-correlation imaging within a single experiment. Particular emphasis was placed on the measures taken to achieve high signal-to-noise ratios, long-term stability of laser beam and setup and accurate temperature control, which are essential for investigating weakly scattering molecular glass formers over timescales up to several days. 

Beyond the experimental implementation, we have outlined a consistent framework for the quantitative analysis of fiber-optical and multispeckle DDLS data in combination with high-frequency DDLS techniques. Reliable comparison of the different techniques requires explicit treatment of coherence effects and partial heterodyne contributions. Once these effects are accounted for, the intensity autocorrelation functions obtained with the different detection schemes can be transformed into mutually consistent electric-field autocorrelation functions. Hence, complementary DDLS techniques allow for the construction of ultra-broadband data sets covering the complete range from microscopic vibrational dynamics to structural relaxation at the glass transition over more than 20 orders of magnitude in time and frequency. In addition, multispeckle detection provides direct access to time-resolved autocorrelation functions without temporal averaging, thereby extending DDLS to the investigation of non-stationary systems such as physically aging molecular glasses.

The methodology presented here provides a basis for future studies of equilibrium and non-equilibrium dynamics in molecular liquids and other weakly scattering soft-matter systems, where high sensitivity, broad dynamic range, and quantitative consistency between complementary experimental techniques are required.

\acknowledgements
Financial support by the Deutsche Forschungsgemeinschaft under grant no.\ 1192/3 is gratefully acknowledged.
We are sincerely grateful to Ernst Rössler, Bayreuth, Germany, for making the Tandem Fabry Perot Interferometer and the double monochromator available to us and for many stimulating discussions. His view on glassy dynamics was inspirational for the experimental developments presented in this paper. Moreover, we are indebted to Tina Hecksher and Jeppe Dyre from the Glass-and-Time group in Roskilde, Denmark, for providing the \emph{Huginn} peltier controller for the presented experiment.

\section*{Conflict of Interest Statement}

The authors have no conflicts to disclose.

\section*{Data availability statement}

The data that support the findings of this study are available from the corresponding author upon reasonable request.

\bibliography{camera_new.bib}
\end{document}